\documentclass[aps,prl,floats,amssymb,twocolumn,letterpaper,groupedaddress,
showpacs,floatfix,twoside]{revtex4}
\usepackage{amssymb,amsmath}

\begin{document}
\title{Interactions Mediated by Surface States: From Pairs
and Trios \\to Adchains and Ordered Overlayers}

\author{Per Hyldgaard}
\email[]{hyldgaar@fy.chalmers.se}
\thanks{Tel: +46 31 772 8422; Fax: +46 31 772 8426}
\affiliation{Department of Applied Physics, Chalmers
University of Technology and G\"{o}teborg University, S-412 96
G\"{o}teborg, Sweden}
\author{T. L. Einstein}
\thanks{Corresponding author}
\email[]{einstein@umd.edu}
\homepage[Webpage: ]{http://www2.physics.umd.edu/~einstein/}
\thanks{Tel. 1-301-405-6147; Fax: 1-301-314-9465 \vspace{5mm}}
\affiliation{Department of Physics, University of Maryland, College Park, Maryland 20742-4111 USA}
\vspace{1mm}
\date{Aug. 6, 2004; {\bf Proc. ICCG-14, Grenoble, Aug. 2004; to be published in J. Crystal Growth (2005)}}

\begin{abstract}
Since metallic surface states on (111) noble metals are
free-electron like, their propagators can be evaluated
analytically.  Since they are well-screened, one can use
simple tight-binding formalism to study their effects.
The needed phase shifts can be extracted
from experiment.  Hence, one can now make quantitative
predictions of these slowly-decaying, oscillatory indirect
interactions.  For the (isotropic!) pair interactions (which decay
as the inverse square of adatom-adatom separation), remarkable
agreement has been obtained with experiments by two groups.
We have extended the formalism to consider the full indirect
(``triple") interaction of 3 adsorbates, which is the sum of the 3
constituent pair interactions plus the non-pairwise ``trio"
contribution, which tends to decay with the 5/2 power of perimeter.
Here, we concentrate on interactions due to ordered overlayers and
to linear defects, relating the latter to the interactions of
(n$\times$1) ordered overlayers and both to the constituent pair
and trio interactions.  We compare with experimental studies of
interactions of adatoms with adchains and of consequent 1D motion
of adatoms trapped between two such parallel chains.  We discuss
implications for step-step interactions (on vicinal surfaces),
with attention to the modification of the surface state itself for
small terrace widths.

Puisque le mod\`ele d'\'electron libres s'applique 
pour les \'etats m\'etalliques de surface (111) des m\'etaux nobles,
les propagateurs peuvent \^etre \'evalu\'es analytiquement. Puisqu'ils sont bien \'ecrant\'es, on peut simplement employer le formalisme  des liaisons fortes pour \'etudier leurs effets. Les d\'ephasages n\'ecessaires peuvent \^etre extraits de l'exp\'erience. Par cons\'equent, on peut maintenant faire des pr\'evisions quantitatives pour ces  interactions indirectes oscillantes
et d\'ecroissant lentement.
Pour  les interactions isotropiques de paires (qui  d\'ecroissent comme l'inverse 
du carr\'e de
la distance entre adatomes), un accord remarquable a \'et\'e obtenu avec des exp\'eriences par deux groupes. Nous avons prolong\'e le formalisme pour consid\'erer l'interaction indirecte  de 3 adsorbants, qui est la somme des 3 interactions constitutives de paires plus la contribution \`a trois corps (trio), qui tend \`a  d\'ecro\^\i tre comme la puissance de 5/2 du p\'erim\^etre. Ici, nous nous concentrons sur les interactions dues aux  couches adsorb\'ees ordonn\'es et les d\'efauts lin\'eaires, reliant ce dernier aux interactions (n$\times$1) des adorbats ordonn\'es et  tous les deux aux interactions constitutives de paires et de trio. Nous comparons avec des \'etudes exp\'erimentales des interactions des adatomes avec des ad-chaines et du mouvement \`a 1D des adatomes emprisonn\'es entre deux telles cha\^ines  parall\`ele. Nous discutons des implications pour les interactions les marches (sur des surfaces  vicinales), en faisant attention \`a la modification de l'\'etat de surface pour de petites largeurs de terrasse.
\end{abstract}

\maketitle
\section{Introduction and Parameters}
\vspace{-4mm}
Metallic surface states, i.e. surface states crossing the Fermi
level, have dramatic consequences that can be explored at the
atomic scale by modern surface probes such as scanning-tunneling
microscopy (STM).  Here, we summarize
our progress to date in understanding the consequences of these
states for nanoscale interactions not only between adsorbed atoms
but also between chains of atoms and other atoms or chains.  This
work is preparatory to extensions to step interactions.  We also
present some helpful tabulations not published previously.

While the simple asymptotic expressions for indirect interactions
are valid only for separations larger than several/many atomic
spacings, the more general expressions are valid for any
surface-lattice separation (and could indeed be generalized to
arbitrary separations by allowing different phase factors for the
interacting species).  On the other hand, for atoms at
nearest-neighbor spacings, particularly for homoepitaxy (or for
adatoms larger than substrate atoms), direct interactions should
also come into play, and are then expected to overwhelm any
indirect effects.  Thus, for example, one must be cautious about
using the formalism below to predict interactions between dimers
or chains and atoms, specific any indirect interactions that
involve a propagator between the two members of the dimer (or
neighboring atoms in a chain).

Table I summarizes key parameters that characterize the relevant
isotropic Shockley surface states found on the (111) facet of the
noble metals Cu and Ag. For both surfaces there now exist
experimental investigations of the long-ranged adsorbate
interactions~\cite{Gerhard,Brune}. The table compares experimental
values of the surface-state band parameters, obtained via
STM~\cite{Jeandupeux}, with our large, sized-converged
calculations by standard first-principles DFT~\cite{DFT,DACAPO};
the agreement is good. The table also shows STM measurements of
the scattering phase shifts $\delta_F\neq 0$
reported~\cite{Gerhard,Brune,Wahetal} for various adsorbates, e.g.
from standing waves in ``quantum corrals" \cite{CroLutEiga}.
Finally the table shows estimates for the Thomas-Fermi screening
length $k_{TF}^{-1}$~\cite{Ashcroft}. The surface-state electron
response arises at much longer length scales, $\lambda_F/2 \gg
k_{TF}^{-1}$, and so will dominate the long-range adsorbate
interaction.

\begin{table}
\caption{Shockley surface-state parameters and Thomas-Fermi
(bulk-screening) wavevectors of the Cu  and Ag (111) surfaces. The
Shockley band is characterized by the effective electron mass
$m_{\rm eff}$, a Fermi energy $\epsilon_F$ (measured relative to
the bottom of the surface-state band), and a corresponding
in-surface Fermi wavevector $q_F=\hbar^{-1}\sqrt{2m_{\rm
eff}\epsilon_F}$ and half wavelength $\lambda_F/2 = \pi/q_F$. The
Thomas-Fermi screening lengths, $k_{TF}^{-1}$, are obtained as in
Ref.~\cite{Ashcroft}. Table adapted from
refs.~\cite{Gerhard,JPCM2000, EPL,HE-SSASS}}
\label{tab:SSpar}
\begin{center}
\begin{tabular}{|l|lc|lc|}
\hline
 &
STM Cu(111)& DFT & STM Ag(111)& DFT\\
\hline
$\epsilon_F$ (eV) & 0.38$^{a}$ & 0.42 & 0.065$^{a}$ & 0.045\\
$m_{\rm eff}/m_e$ & 0.44$^{a}$ & 0.38 & 0.40$^{a}$ & ---\\
$q_F$ (\AA$^{-1}$) & 0.21$^{a}$ & 0.20 & 0.083$^{a}$ & ---\\
$\lambda_F/2$ (\AA) & 15.0$^{a}$ & 15.5 & 37.9$^{a}$ & ---\\
$k_{\rm TF}^{-1}$ (\AA) &\multicolumn{2}{c|}{0.552} &
\multicolumn{2}{c|}{0.588}\\
\hline S adsorbate & $\delta_F=\pm \pi/2$$^{b}$ & --- &
--- & --- \\
Cu adsorbate & $\delta_F=\pm \pi/2$$^{c,d}$ & --- &
--- & --- \\
Co adsorbate & $\delta_F=\pm \pi/2$$^{d}$ & --- &
$\delta_F=\pi/3$$^{d}$ & --- \\
\hline \multicolumn{5}{l}{${}^a$Ref.~\cite{Jeandupeux} \qquad
${}^b$Ref.~\cite{Wahetal} \qquad
${}^c$Ref.~\cite{Gerhard} \qquad ${}^d$Ref.~\cite{Brune}}\\
\end{tabular}
\end{center}
\end{table}

\section{Pair interactions}
\vspace{-4mm}
The interaction between adsorbates on a metal surface can involve
an elastic, an electrostatic, and an indirect coupling through
electronic states of the substrate. The long history of
theoretical investigation of indirect adsorbate interactions dates
back nearly four decades~\cite{Grimley}; the history of this
oscillatory, long range interaction has been amply
documented~\cite{EinRev}. Lau and Kohn~\cite{LauKohn} pointed out
that the range of the interaction increases dramatically when the
mediation is by a surface rather than a bulk states. Recent theory
work~\cite{Gerhard,JPCM2000} applied these ideas to the
above-mentioned isotropic surface-state bands to find the
pair-interaction~\cite{Gerhard,JPCM2000}:
\begin{eqnarray}
\Delta E_{\rm pair}(d;\delta_F) =  \frac{2}{\pi} \, {\rm Im}
\int_0^{\epsilon_F}\! \! d\epsilon \, \ln \left(1 -
[t_0(\epsilon;\delta_F)
g_0(qd)]^2 \right) \\
\sim  \Delta E_{\rm pair}^{\rm
asym}(d;\delta_F) =
 -\epsilon_{\rm F}
\left(\frac{2\sin(\delta_{F})}{\pi}\right)^{2}
\frac{\sin(2q_{\rm F}d + 2\delta_F)}
{(q_{\rm F}d)^{2}}.
\label{eq:2DEpairRes}
\end{eqnarray}
The simple analytic expression holds at asymptotic separation
$d>\lambda_F/2$.  The effective T-matrix
$t_0(\epsilon;\delta_F)=-(2\hbar/m_{\rm eff})
\sin(\delta_0(\epsilon))\exp(i\delta_0(\epsilon))$, is determined
by the $s$-wave phase shift $\delta_0(\epsilon)$ with 
the boundary condition $\delta_0(\epsilon_F)=
\delta_F$.  The surface propagator $g_0(x)$ becomes basically the
cylindrical Hankel function of the first kind ($H_0^{(1)})$:
\begin{equation}
g_0(x)  =  i\frac{m_{\rm eff}}{2\hbar}H_0^{(1)}(x)
          \sim  i\frac{m_{\rm eff}}{\hbar}\frac{\exp(ix-i\pi/4)}
          {\sqrt{2\pi x}}, \quad x
\rightarrow \infty. \label{eq:g2DDef}
\end{equation}
To obtain the simple asymptotic
expressions, $tg$ must be small enough so that $\ln [1 - \ldots]$
can be expanded to leading order and $x$ must be large enough to
replace $H_0^{(1)}(x)$ by an outgoing circular wave.

Subsequent STM measurements of Cu and Co adsorbate dynamics on
Cu(111) and Ag(111)~\cite{Gerhard,Brune} have verified that the
interaction has period $\lambda_F/2 = \pi/q_F$ and the quadratic
decay of the envelope with separation, both without adjustable
parameters.  Accounting for the overall magnitude requires
insight into inelastic losses to bulk states.

\section{Trio interactions}
\vspace{-4mm}
Study of the interaction of three
adsorbates~\cite{EPL,HE-SSASS} serves as a bridge from pair
interactions to multi-adsorbate interactions in clusters.
The three adsorbates are taken to bond
to substrate positions $i=1,2,3$. The triple-adsorbate cluster
adsorption energy is calculated~\cite{EPL} by combining a formal
expansion~\cite{EinRev,tle-trio} of the adsorbate-cluster energy
with scattering theory~\cite{JPCM2000}:
\begin{eqnarray}
\lefteqn{\Delta E_{\rm triple}(d_{12},d_{23},d_{31};\delta_F)} \nonumber \\
&&\hspace{-3mm} \equiv \sum_{i > j=1}^3 \Delta E_{\rm pair}(d_{ij};\delta_F) +\Delta
E_{\rm trio}(d_{12},d_{23},d_{31};\delta_F) \\
&&\hspace{-3mm}=  \frac{2}{\pi} \, {\rm Im} \int_0^{\epsilon_F}\! \! \! d\epsilon \, \ln
\left[1 -\! \left(K_{12}^2\! +\! K_{23}^2\! +\! K_{31}^2\right)
-\! 2K_{12}K_{23}K_{31}\right] , \nonumber
\label{eq:2DEtripledef}
\end{eqnarray}
where $K_{ij}$ is shorthand for
$t_0(\epsilon;\delta_F)g_0(qd_{ij})$. This triple-cluster
interaction includes a new trio contribution $\Delta E_{\rm trio}$
which arises from constructive interference of electrons which
traverse the entire cluster parameter
$d_{123}=d_{12}+d_{23}+d_{31}$.  In the asymptotic limit, $d_{123}
> 3 \lambda_F$,  we obtain the analytical
result~\cite{EPL,HE-SSASS}:

\begin{eqnarray}
\lefteqn{\Delta
E_{\rm{trio}}(d_{12},\! d_{23},\! d_{31};\! \delta_F)
  \simeq - \frac{4}{\pi} \,
{\rm Im}  \int_0^{\epsilon_F}\! \! d\epsilon
\, K_{12}K_{23}K_{31}} \\
&&\hspace{-2mm} \sim -\epsilon_{\rm F}
\sin^{3}(\delta_F)
\left(\frac{16\sqrt{2}}{\pi^{5/2}}\right)
\gamma_{123}\frac{\sin(q_{\rm F}d_{123}
+ 3\delta_F -\frac{3}{4}\pi)}{(q_{\rm F}d_{123})^{5/2}}. \nonumber
\label{eq:2DEtrioRes}
\end{eqnarray}
For completely absorbing scatterers the trio
interaction result~(\ref{eq:2DEtrioRes}) is reduced by a factor of
1/8 (see Ref.~\cite{EPL}).  Since the scattering is taken to be s-wave, the trio
interaction depends overwhelmingly on the perimeter $d_{123}$ and
is insensitive to the shape: the geometrical prefactor
$\gamma_{123} \equiv \sqrt{d_{123}^3/d_{12}d_{23}d_{31}}$ varies little except for
highly distorted arrangements~\cite{EPL,HE-SSASS}.
Also~\cite{HE-SSASS}, trio interactions can affect the barriers of
atoms approaching growing clusters, an issue of recent theoretical
study \cite{Fichthorn}.

Our results are summarized in Table 2.  We emphasize that our
calculations are non-perturbative, resulting in the physically-important
phase shift $\delta_F$ absent in
perturbative approaches (e.g. Ref.~\cite{LauKohn}).
Since $\delta_F$
can differ for various adatom-substrate combinations (cf.~Table 1), one
can in principle select a system that will have a minimum at an
arbitrary lattice spacing.

\begin{table}\caption{Comparison of indirect interactions on surfaces mediated
by [metallic] surface and bulk states and as well as bulk
interactions (mediated by bulk states).  The pair and trio decays
refer to the envelope of the oscillatory interaction in the
asymptotic regime.}
\label{tab:bulksurf}
\begin{center}
\begin{tabular}{|l|c|c|c|}
\hline  &Surf. via surface& Surf. via bulk & Bulk \\
\hline \hline
 $\lambda_F /2$ &  $\sim$
15.0\AA \
[Cu(111)] & $\sim$ 2.3\AA \ [Cu] & $\sim$ 2.3\AA \ [Cu] \\ \hline
Dispersion & Isotropic &
Anisotropic  &  Anisotropic \\
&$\epsilon \approx (\hbar
k_{\parallel})^2/2m^{\ast}$&$\epsilon_n({\rm \bf k}_{\parallel})$&
$\epsilon_n({\rm \bf k})$\\ \hline
Compu- &Simple:
para-   & Messy: multi- & Messy\\
\hspace{1mm}tation&bolic 2D band&ple 3D bands& \\ \hline
Pair decay & $\propto d^{-2}$  &
$\propto d^{-5}$ & $\propto d^{-3}$  \\
&$\Rightarrow$  observable&$\Rightarrow$ insignificant& RKKY \\ \hline
Trio decay & $\propto d^{-5/2}$ & $\propto d^{-7}$& $\propto d^{-4}$\\
 \hline
\end{tabular}
\end{center}
\end{table}
While some evidence exists that the pair interaction alone
is inadequate at non-asymptotic separations, there has not yet
been a comparable experimental confirmation; trio interactions
between adatoms and dimers are likely to be dwarfed by
direct-interaction effects in the dimer, but other effects can be
envisioned.

The preceding process can be extended to compute interactions
between 4, 5, and more adatoms.  The formalism for bulk
impurities, readily convertible to surfaces, was worked out by
Harrison~\cite{Harrison}.  Alternatively, one can consider the
interaction energy of superlattices of
adsorbates~\cite{EinRev,multi}.  In this way, one can relate the integrand
for $\Delta E$ of a fractional overlayer to that of a full monolayer.  However,
the simple expression for a full monolayer given in Ref.~\cite{multi} involves 
``tricks" related to the simple model employed that are subtle to generalize.

\section{Interactions with chains}
\vspace{-4mm}
By viewing a chain as the sum of its constituent atoms, one can
readily add up these interactions~\cite{HE2} to show

\begin{equation}
  \Delta E_{\rm chain-atom}^{\rm asym}(\ell) \propto
 -\epsilon_{\rm F}\,\,
\sin^2(\delta_{F})\, \frac{\sin(2q_{\rm F}\ell + 2\delta_F + \pi/4)}
{(q_{\rm F}\ell)^{3/2}}, \label{eq:chainatom}
\end{equation}
\noindent where $\ell$ is the distance from the atom to the chain.
The remarkable 3/2 power law was recognized over a decade
ago~\cite{RedZang}. 
A similar result should arise from consideration of the interaction energy of an
(n$\times$1) array of adatoms (but cf.~warning at the end of section 3).

Inserting parameters for Cu(111) into Eqn.~(\ref{eq:chainatom}),
we find minima when $\ell$ is 9, 24, 39, and 54 \AA.  In counting
the occurrences of atoms between 20 and 30 \AA \ from a chain,
Repp~\cite{ReppPhD} did indeed find the behavior of
Eqn.~(\ref{eq:chainatom}).  {\em The chain-chain interaction 
has the same form as Eqn.~(\ref{eq:chainatom})} since the second
chain can (also) be viewed as the sum of individual atoms, each of
which have this interaction.

An atom between two parallel chains will experience a 1D
corrugation potential parallel to the chains.  Repp constructed
such a situation for Cu atoms on Cu(111) with atomic
manipulation~\cite{ReppPhD} and produced STM movies of atoms
wandering along the trough. Since the chains are of finite length,
the well depth decreases near the ends of the chain.  Hence, the
atom is trapped in this furrow.  If the chains are far enough
apart, there are multiple furrows.  Repp~\cite{ReppPhD} observed
two atoms, in furrows nanometers apart, moving back and forth
individually.

One can imagine extensions of these ideas such as producing
gridworks of chains with a regular set of traps for atoms or a
maze of walls through which atoms might move as stupid rats.
Computing the corresponding potential surface is then a fairly
well-defined task.

\section{Complications in going from chains to steps}
\vspace{-4mm}
Surface states are not so robust as bulk states, so one cannot blithely
view them as unaffected by the adsorption process.  Baumberger et al.~\cite{BG02}
show that, on vicinal Cu(111), the surface state is shifted up (and so $q_F$ reduced)
as the terrace width $\ell$ decreases.  However, when the steps are decorated with CO, the energy shift becomes downward with decreasing $\ell$!

Furthermore, Ortega et al.~\cite{OH00} find that when $\ell$ decreases 
sufficiently (in particular, when the misorientation of Cu(111) increases beyond 
~7$^{\circ}$), the surface state is no longer that of the (111) facet but is determined
by the vicinal surface itself.  The periodic potential of the steps then opens a gap in 
the parabolic band structure.

Both these arguments assume implicitly that the steps are straight and uniformly spaced, neither of which are generally true.  It is not clear how the meandering of steps or the fluctuations in $\ell$ alter these results or, conversely, how the interactions affect the meandering and distribution of the steps.  (In concise words, are the steps ``actors or spectators?"~\cite{G04}.) 

The existence of slowly-decaying oscillatory interactions should have profound implications for the distribution of terrace widths $P(\ell)$.  In general the dominant interaction between steps comes from entropic and elastic repulsions, both of which vary as $\ell^{-2}$.  As a consequence $P(\ell)$ has a ``universal" form depending only on the ratio $\ell/\langle \ell \rangle$ (and the strength of the $\ell^{-2}$ repulsion) but not on the mean spacing $\langle \ell \rangle$, i.e., not on the misorientation.  With surface states, this scaling breaks down, as has been observed experimentally~\cite{PR94}.

Furthermore, the oscillatory interaction introduces a new length scale $\lambda_F$.  Thus, the equilibrium crystal shape, which is expected to be independent of crystal size, would seem to acquire some size-dependent behavior, at least for small crystallites.  Since the $\ell^{-3/2}$ decay of the envelope is slower than the $\ell^{-2}$ of the pure repulsion, it is not clear what changes arise in the Pokrovsky-Talapov~\cite{PT} ``critical behavior" of the curved regions near the edges of facets.

\section{Conclusions}
\vspace{-4mm}
\noindent In summary, we present both an asymptotic evaluation and
an exact model calculation for adsorbate interaction energies
mediated by an isotropic Shockley surface-state band, as found on
noble-metal (111) surfaces.  While this interaction is primarily
the sum of pair interactions, there can be significant trio
corrections. Such interactions can play a role in the
low-temperature adsorbate assembly~\cite{EinRev,Fichthorn,Alex},
and efforts are being made to investigate them
directly~\cite{Brune,MorgensternPRL}. We can on this basis evaluate
the interaction between a chain of adatoms and another chain
and/or other adatoms. Novel nanostructures can be imagined and
actually contructed~\cite{ReppPhD} by skilled experimentalists.

As noted, the slowly-decaying oscillatory interactions should affect a broad range of phenomena and should apply to any situations in which defects create localized perturbations on surfaces with surface states, e.g., magnetic interactions.  Thus, the exchange coupling should oscillate with the same period $\lambda_F/2$ as the adatom-adatom interaction; however, there is no {\it a priori} reason to expect that the phase shift $\delta_F$ will be the same.  Thus, one can imagine a rich phase diagram.

In subsequent papers we will present a detailed
investigation~\cite{HE2} of the surface-state-derived interactions
associated with chains and nanostructures.  We will also produce a
careful and thorough analysis and assessment of the assumptions
involved in our approach~\cite{HPE}, with comments about
extensions to systems in which, for example, rapid screening of
the adsorption bond is questionable.

\section*{Acknowledgments}
\vspace{-4mm}
This work was supported by (PH) W.~\& M.~Lundgrens Foundation,
Swedish Foundation for Strategic Research (SSF) through consortium
ATOMICS, and by (TLE) NSF through MRSEC Grant DMR 00-80008 and
Grant EEC-0085604. We are grateful to many of the cited
experimentalists for enlightening interchanges.  We thank Jascha
Repp in particular for unpublished information, movies, and
results from his dissertation about chains and adatoms and/or
other chains.
\vspace{-4mm}


\end{document}